\documentclass{article}
\usepackage{spconf,amsmath,amssymb,graphicx,booktabs,microtype}
\usepackage[hidelinks]{hyperref}
\graphicspath{{figures/}}

\title{A CORPUS-ONLY AUC DIAGNOSTIC FOR POOLED SPEECH-BASED DEPRESSION DETECTION}

\name{Cezary Polak$^{1}$, Milosz Dudek$^{1,2}$, Daria Hemmerling$^{1,2}$}
\address{$^{1}$SoftServe, Krakow, Poland
\quad
         $^{2}$AGH University of Krakow, Krakow, Poland }

\begin{document}
\pagestyle{empty}
\maketitle

\begin{abstract}
When clinical speech corpora are pooled for evaluation, corpus membership can itself
become predictive if class prevalences differ across sources. We derive the AUC of an
oracle predictor that uses only corpus identity and empirical prevalence, a diagnostic
computable directly from corpus composition. Across five depression corpora in three
languages (590 speakers), the corpus-only AUC is 0.697, and corpus identity is predicted
from six acoustic and self-supervised representations with 97.5--99.8\,\% accuracy.
Permuting depression labels within each corpus removes label structure while preserving
prevalences; the residual pooled-minus-LOCO discrepancy then increases across three
nested configurations in the same order as the diagnostic. Pooled 5-fold
cross-validation also outscores corpus-held-out evaluation across all six
representations. Prevalence equalisation substantially reduces the discrepancy for
eGeMAPS and WavLM-Base, whereas a sample-size-matched control does not. Corpus-specific
base rates therefore create a measurable shortcut opportunity in pooled clinical-speech
evaluation, motivating corpus-aware diagnostics alongside pooled performance.
\end{abstract}


\section{Introduction}
\label{sec:intro}

Pooling small clinical-speech corpora is attractive because individual datasets are
often too small to support stable model comparison. Yet a pooled split changes the
prediction problem: training and test folds may contain different speakers while still
sharing corpus-specific acquisition conditions, language, elicitation, recruitment,
and label distributions. A model can exploit source information even under strictly speaker-disjoint
cross-validation.

\begin{sloppypar}
Cross-corpus degradation itself is well established. Schuller et
al.~\cite{schuller2010crosscorpus} showed large intra- to inter-corpus losses in speech
emotion recognition, while Mitra et al.~\cite{mitra2015crosscorpus} reported only
partial and elicitation-dependent transfer for speech-based depression prediction.
More generally, shortcut learning describes models exploiting predictive regularities
that are stable in a benchmark but are not the intended signal
\cite{geirhos2020shortcut}. In depression recognition, Burdisso et
al.~\cite{burdisso2024daic} demonstrated a concrete shortcut in DAIC-WOZ, where
interviewer-related structure can yield deceptively strong performance.
\end{sloppypar}

Here we isolate a simpler population-level mechanism. If corpora are acoustically
identifiable and have different positive prevalences, corpus membership alone induces a
ranking signal: a subject drawn from a high-prevalence corpus can receive a higher score
than one from a low-prevalence corpus before any within-corpus depression cue is used.
We ask how much discrimination is available from corpus identity and
corpus-specific base rates alone, and whether this channel is accessible to
standard speech representations.

\noindent\textbf{Contributions.}
(i)~We derive a closed-form corpus-only AUC that quantifies the discrimination available
from corpus membership and corpus-specific class prevalence alone.
(ii)~We audit five depression speech corpora using six acoustic and self-supervised
representations, combining corpus-identification probes with within-corpus label
permutation to test whether the base-rate channel is accessible to the models.
(iii)~We examine the consequences for evaluation through pooled and corpus-held-out
protocols, and use prevalence-equalisation and matched-size controls to test whether the
observed discrepancy tracks corpus-specific base rates.

\section{Data and method}
\label{sec:data}

\begin{table}[t]
\caption{Corpora included in the audit. All six representations are available for the
same 590 speakers.}
\label{tab:corpora}
\centering\small
\setlength{\tabcolsep}{4pt}
\begin{tabular}{@{}llrrr@{}}
\toprule
Lang & Corpus (elicitation) & $N$ & Pos & Prev \\
\midrule
EN & DAIC-WOZ~\cite{gratch2014distress} (interview)  & 141 & 41  & 29\,\% \\
ZH & EATD~\cite{shen2022automatic} (prompts)         & 162 & 30  & 19\,\% \\
ZH & MPDD-Elderly~\cite{fu2025mpdd} (interview)      & 87  & 11  & 13\,\% \\
ZH & MPDD-Young~\cite{fu2025mpdd} (read + intro)     & 88  & 10  & 11\,\% \\
IT & Androids-Read~\cite{tao2023androids} (passage)  & 112 & 58  & 52\,\% \\
\midrule
   & \textbf{Total}                                  & \textbf{590} & \textbf{150} & \textbf{25\,\%} \\
\bottomrule
\end{tabular}
\end{table}

\noindent\textbf{Corpora and labels.} Table~\ref{tab:corpora} summarises the five
datasets. Binary targets follow the clinically motivated criterion available for each
corpus: PHQ-8/PHQ-9 $\geq 10$ \cite{kroenke2009phq8}, Zung SDS $\geq 53$, and clinical
diagnosis for Androids. MPDD is re-thresholded from its original mild-and-above
definition to obtain a more comparable binary depression-related endpoint. These
criteria provide operationally aligned binary labels across instruments. The resulting prevalences range from
11--52\,\%, creating the corpus-level base-rate variation analysed below.

\noindent\textbf{Preprocessing and representations.} Audio is resampled to 16\,kHz mono
and RMS-normalised to $-23$\,dBFS. For the self-supervised encoders, recordings longer
than 30\,s are divided into 30-s windows; frame-level hidden states are mean-pooled to a
window embedding and window embeddings are averaged. eGeMAPS~v02 (88-d)
\cite{eyben2016gemaps} is extracted with openSMILE \cite{eyben2010opensmile} as
utterance-level functionals. When a subject contributes multiple clips, their vectors are
averaged to obtain one subject-level representation. DAIC-WOZ additionally uses
transcript-based participant-only audio before feature extraction
\cite{burdisso2024daic}. We evaluate WavLM-Base and WavLM-Base+
\cite{chen2022wavlm}, HuBERT \cite{hsu2021hubert}, XLS-R \cite{babu2022xlsr}, and
mHuBERT-147 \cite{boito2024mhubert}; all SSL encoders are frozen.

\noindent\textbf{Classifier.} Depression-prediction experiments use
$L_2$-regularised logistic regression on frozen representations, with feature
standardisation fitted on the training fold only. 
Pooled cross-validation uses global class-balanced weights. In the current
corpus-held-out implementation, training weights assign equal total mass to each
training corpus and equal mass to the classes within each corpus.

\noindent\textbf{Protocols.} \emph{Pooled 5-fold cross-validation} is the evaluation
protocol under audit. \emph{Leave-one-corpus-out} (LOCO) holds out one complete corpus
at a time and is used as a cross-corpus generalisation stress test. Because the current
pooled and LOCO estimators use different weighting schemes, their difference is not
interpreted as a decomposition of pooled AUC into clinical and corpus-specific
components. The mechanistic analysis instead relies on within-corpus label permutation
and the prevalence interventions described below.

\section{Corpus identity and cross-corpus evaluation}
\label{sec:gap}

A multinomial logistic-regression probe predicts \emph{corpus identity} from the
representations with 97.5--99.8\,\% stratified 5-fold accuracy, compared with a
27.5\,\% majority baseline. The corpus-only diagnostic derived in
Sec.~\ref{sec:closedform} gives AUC 0.697 without using the speech signal.

Corpus identification is not driven by a single dominant source. In the final audit,
removing the largest corpus (EATD) leaves 428 speakers across four corpora, and corpus
identity remains recoverable at 99.5--100\,\% across the six representations.

\begin{table}[t]
\centering
\caption{Pooled 5-fold cross-validation and corpus-held-out (LOCO) performance across
representations, averaged over 30 fold assignments. The final column reports the
difference between the two protocols. }
\label{tab:gap}
\begin{tabular}{@{}lrrr@{}}
\toprule
Representation & pooled & LOCO & $\Delta$ \\
\midrule
eGeMAPS       & 0.629 & 0.475 & $+0.154$ \\
WavLM-Base    & 0.675 & 0.467 & $+0.208$ \\
WavLM-Base+   & 0.680 & 0.472 & $+0.209$ \\
HuBERT        & 0.665 & 0.456 & $+0.208$ \\
XLS-R         & 0.725 & 0.434 & $\mathbf{+0.291}$ \\
mHuBERT-147   & 0.681 & 0.461 & $+0.220$ \\
\bottomrule
\end{tabular}
\end{table}

Table~\ref{tab:gap} shows the discrepancy for both hand-crafted eGeMAPS and all
tested self-supervised representation families. The permutation and prevalence
analyses below test the base-rate mechanism directly.

\section{Corpus-only AUC diagnostic}

\subsection{Closed-form corpus-only AUC}
\label{sec:closedform}

For corpora $k$ with $n_k$ subjects at prevalence $p_k$, writing $P_k = n_k p_k$ and
$N_k = n_k(1-p_k)$, consider an oracle predictor that knows only corpus identity and
assigns every subject from corpus $k$ the score $p_k$. Its AUC is
\begin{equation}
\begin{aligned}
\mathrm{AUC}_{\mathrm{corpus}}
&=\frac{2\!\!\displaystyle\sum_{j,k:\,p_j>p_k}\!\!P_jN_k
      +\!\!\displaystyle\sum_{j,k:\,p_j=p_k}\!\!P_jN_k}{2PN},\\[-1mm]
P&=\sum_j P_j,\qquad N=\sum_k N_k .
\end{aligned}
\label{eq:corpus_auc}
\end{equation}
Both sums run over all ordered corpus pairs $(j,k)$, including $j=k$, so the
equal-prevalence term includes within-corpus positive-negative pairs. We call
Eq.~(\ref{eq:corpus_auc}) the \emph{corpus-only AUC}. It quantifies the discrimination
available from corpus identity and empirical corpus prevalences alone; it is not an
upper bound on model AUC and should not be subtracted from model performance. For the five corpora in Table~\ref{tab:corpora},
$\mathrm{AUC}_{\mathrm{corpus}}=\mathbf{0.697}$.

\subsection{Within-corpus label permutation}
\label{sec:validate}




To test whether the corpus base-rate channel is accessible through the speech
representations, we permute labels \emph{within each corpus} and rerun both evaluation
protocols. The permutation destroys systematic within-corpus association between speech
and depression label while preserving corpus sizes and corpus-specific prevalences. We report the
residual pooled-minus-LOCO difference after permutation: the part of the evaluation
discrepancy that survives once the within-corpus label signal is destroyed.

Using the four SSL representations available in all three nested configurations, the
mean residual discrepancy increases in the same order as the analytic diagnostic:
$+0.003$ at $\mathrm{AUC}_{\mathrm{corpus}}=0.567$ for the three Mandarin corpora
($n=337$), $+0.023$ at 0.621 after adding DAIC-WOZ ($n=478$), and $+0.060$ at 0.697
after adding Androids-Read ($n=590$). The final leakage audit used 30 fold assignments
and 200 within-corpus label permutations. With only three independent corpus configurations available, we report the ordering
and do not fit a correlation.

\subsection{Prevalence equalisation}
\label{sec:equalise}

Eq.~(\ref{eq:corpus_auc}) predicts that reducing prevalence differences between corpora
should reduce $\mathrm{AUC}_{\mathrm{corpus}}$ and, if the base-rate channel contributes
to pooled performance, reduce the pooled-versus-held-out discrepancy. Subsampling each
corpus to a common prevalence of 0.19 reduces
$\mathrm{AUC}_{\mathrm{corpus}}$ from 0.697 to 0.501
(Table~\ref{tab:interventions}). The discrepancy changes from $+0.113$ to $-0.032$
[$-0.106$, $+0.042$] for eGeMAPS and from $+0.206$ to $+0.023$
[$-0.067$, $+0.115$] for WavLM-Base. Both bootstrap intervals include zero, so
equalisation substantially reduces the discrepancy without demonstrating that it is
eliminated.

As a training-only alternative, we also apply prevalence importance weights. For corpus
$k$ with prevalence $p_k$ and target prevalence $p^\star$, cases receive weight
$p^\star/p_k$ and controls receive $(1-p^\star)/(1-p_k)$. The ``+ equal mass'' variant
additionally normalises the weights so that every training corpus carries equal total
mass.

\begin{table}[t]
\caption{Independent control analysis for eGeMAPS and WavLM-Base. Values in the final
two columns are pooled-minus-LOCO differences. For the full, mass-matched, and
prevalence-equalised conditions, both estimators are recomputed on the subjects retained
in that condition; importance-weighting conditions retain all subjects.}
\label{tab:interventions}
\centering\small
\setlength{\tabcolsep}{2pt}
\begin{tabular}{@{}lrrcc@{}}
\toprule
 & $n$ & corpus AUC & eGeMAPS & WavLM-Base \\
\midrule
full data          & 590 & 0.697 & $+0.113$ & $+0.206$ \\
mass-only          & 455 & 0.697 & $+0.102$ & $+0.209$ \\
\textbf{equalised} & 454 & \textbf{0.501} & $\mathbf{-0.032}$ & $\mathbf{+0.023}$ \\
weights            & 590 & 0.697 & $+0.052$ & $+0.194$ \\
\;\;+ equal mass   & 590 & 0.697 & $+0.058$ & $+0.193$ \\
\bottomrule
\end{tabular}
\end{table}

\noindent\textbf{Matched-size control.} Prevalence equalisation necessarily removes
subjects, so part of the reduction could arise from the smaller sample alone. We draw a
proportional subsample of approximately the same size while preserving the original
corpus prevalences. This changes the discrepancy by only $-0.011$ for
eGeMAPS and $+0.003$ for WavLM-Base, compared with reductions of $0.145$ and $0.183$
after prevalence equalisation.

\begin{sloppypar}
A calibrated dose--response relationship between $\mathrm{AUC}_{\mathrm{corpus}}$ and
the residual discrepancy would require additional independent corpora, not further
resampling of the same datasets.
\end{sloppypar}

\section{Discussion and limitations}

Speaker-disjoint pooled cross-validation is not by itself evidence of cross-corpus
generalisation. Corpus composition should be reported together with a simple
corpus-only diagnostic such as Eq.~(\ref{eq:corpus_auc}) when class prevalences differ
substantially between sources. Corpus-held-out evaluation addresses a different
question and should be reported when claims concern generalisation to previously unseen
data sources.

Prevalence equalisation is used here to probe the mechanism; we do not recommend
discarding training data in practice. It reduces the corpus-label association while
retaining the original representations for the selected participants. The matched-size
control produces a much smaller change, supporting a substantial role for prevalence
heterogeneity in the observed discrepancy. Simple importance weighting is
representation-dependent: it reduces the discrepancy for eGeMAPS but produces only a
small change for WavLM-Base, so it does not generalise as a remedy.


\noindent\textbf{Limitations.}
A high corpus-only AUC shows that corpus membership contains label information; how
much a particular classifier exploits it depends on the representation and learning
procedure. Corpus identity also aggregates several factors, including language,
elicitation task, recording conditions, recruitment, and labeling protocol, which
cannot be disentangled with only five corpora. Our experiments use frozen
representations and simple classifier heads, so fine-tuned systems may behave
differently. English and Italian are represented by one corpus each, leaving language
and corpus partially aliased, and the two Mandarin MPDD cohorts contain only 11 and 10
positive subjects. Finally, the binary endpoints are derived from different clinical
instruments and should not be interpreted as clinically interchangeable.

\section{Conclusion}

We introduced a corpus-only AUC diagnostic for quantifying the discrimination available
from corpus membership and corpus-specific prevalence in pooled clinical-speech
evaluation. Across five depression corpora, corpus identity was highly recoverable from
all tested representations, while controlled label-permutation and prevalence
experiments showed that this base-rate channel is empirically accessible. The diagnostic
provides a warning signal for pooled evaluations in which corpus and label
distributions are coupled. We recommend
reporting corpus composition and corpus-held-out performance whenever multi-corpus
results are used to support generalisation claims.

\section{Compliance with ethical standards}

This research study was conducted retrospectively using human subject data made
available by the providers of five existing corpora, each used within its own access
agreement. No new data were collected and no participant was contacted; all corpora
were de-identified by their providers, and ethical approval was not required for this
secondary analysis. Reported AUCs are screening signals evaluated retrospectively and
support no clinical use. No funding was received for conducting this study, and the
authors have no relevant financial or non-financial interests to disclose.

\vfill\pagebreak
\bibliographystyle{IEEEbib}
\bibliography{refs}

\end{document}